\documentclass[twocolumn,floats,floatfix,amssymb,nofootinbib,prd,superscriptaddress,aps]{revtex4-2}
\usepackage{graphicx,amssymb,amsmath,amsthm,amsfonts,epsfig}
\usepackage[linktocpage,colorlinks=true,citecolor=OliveGreen,linkcolor=Maroon,urlcolor=Maroon]{hyperref}
\usepackage[usenames]{color}
\usepackage{epstopdf}
\usepackage{bm}
\usepackage{dcolumn}
\usepackage[utf8]{inputenc}
\usepackage{latexsym}
\usepackage{rotating}
\usepackage{hyperref}
\usepackage{tabularx}
\usepackage{braket}
\usepackage{color}
\usepackage{esvect}
\usepackage{longtable}
\usepackage{enumerate}
\usepackage{footmisc}
\usepackage{array}
\usepackage{tensor}
\usepackage{mathtools}
\usepackage{url}
\usepackage[dvipsnames]{xcolor}
\usepackage{comment}
\usepackage{multirow}
\usepackage{nicematrix}
\usepackage{mathrsfs}
\renewcommand{\vec}[1]{\boldsymbol{#1}}

\newcommand{\ben}{\begin{enumerate}}
\newcommand{\een}{\end{enumerate}}

\def\be{\begin{equation}}
\def\ee{\end{equation}}
\def\bea{\begin{eqnarray}}
\def\eea{\end{eqnarray}}
\def\beq{\begin{eqnarray}}
\def\eeq{\end{eqnarray}}
\newcommand{\ba}{\begin{align}}
\newcommand{\ea}{\end{align}}
\def\ba{\bar{a}}

\newcommand{\scri}{\mathscr{I}}

\definecolor{cornellGreen}{HTML}{6EB43F}
\definecolor{cornellRed}{HTML}{B31B1B}

\begin{document}
\title{Gravitational-wave tails and memory effect for mergers in astrophysical environments}
\author{Qassim Alnasheet}
\affiliation{Center of Gravity, Niels Bohr Institute, Blegdamsvej 17, 2100 Copenhagen, Denmark}
\author{Vitor Cardoso}
\affiliation{Center of Gravity, Niels Bohr Institute, Blegdamsvej 17, 2100 Copenhagen, Denmark}
\affiliation{CENTRA, Departamento de F\'{\i}sica, Instituto Superior T\'ecnico -- IST, Universidade de Lisboa -- UL,
Avenida Rovisco Pais 1, 1049 Lisboa, Portugal}
\author{Francisco Duque}
\affiliation{Max Planck Institute for Gravitational Physics (Albert Einstein Institute) 
Am Mühlenberg 1, D-14476 Potsdam, Germany}
\author{Rodrigo Panosso Macedo}
\affiliation{Center of Gravity, Niels Bohr Institute, Blegdamsvej 17, 2100 Copenhagen, Denmark}
\begin{abstract}
Gravitational waves from the coalescence of compact objects carry information about their dynamics and the spacetime in regions where they are evolving. In particular, late-time tails and memory effects after the merger are two low-frequency phenomena, not detectable by current instruments, but which can be observed by future detectors. Their low-frequency nature could, in principle, make them more sensitive to larger-scale structures at galactic length scales. We show that indeed there are transient features, such as amplitude changes, in both tails and (linear) memory when the merger occurs while immersed in an astrophysical environment. For realistic galaxies, the environment's compactness is small enough that the effect is strongly suppressed, but these effects could become relevant for mergers occurring in regions with matter overdensities, like the ones recently observed numerically for wave dark matter. On the other hand, the memory (the difference between the amplitude asymptotically early and late) and asymptotically late decay are independent on the properties of the environment.
\end{abstract}
\maketitle
%
\section{Introduction}

In a merger of two compact objects, e.g. black holes (BH), the highest energies that are probed are directly related to the smallest spatial scales, namely their sizes, and imprinted on the high-frequency content of the signal, like the \textit{ringdown} where the remnant object relaxes to a quasi-stationary endstate through a superposition of quasinormal modes~\cite{Cardoso:2019rvt,Berti:2009kk,Berti:2025hly}. Yet, the post-merger signal also contains \textit{low-frequency} signatures associated with the interaction gravitational waves (GW) emitted by the binary have on the spacetime as they propagate.

An example are hereditary late-time tails, which arise due to the back-scattering  of propagating GWs on the non-zero spacetime curvature~\cite{Price:1971fb,Gundlach:1993tp,Ching:1995tj,Blanchet:1994ez,Cardoso:2024jme}. At a formal level, tails were first understood in terms of a zero-frequency branch-cut in the Green's function propagating linear perturbations to a background spacetime~\cite{Leaver:1986gd, Ching:1995tj}. This is responsible for a polynomial decay at future null infinity as $u^{-\ell-2}$, with $u$ the retarded time and $\ell$ the multipole, that dominates the post-merger signal after the ringdown. However, this power-law is an asymptotic value that is actually approached via a superposition of other power-laws at intermediate times~\cite{Zenginoglu:2012us}. The excitation of these intermediate tails was recently understood to be \textit{source-driven}, and thus highly sensitive to the binary's dynamics~\cite{DeAmicis:2024eoy, Cardoso:2024jme, Islam:2024vro}, in particular its eccentricity~\cite{DeAmicis:2024not}. Highly eccentric coalescences can enhance the tail amplitude by more than 4 orders of magnitude with respect to the circular case. Remarkably, all these features have also recently been confirmed by Numerical Relativity~\cite{DeAmicis:2024eoy,Ma:2024hzq} in evolutions of the full, non-linear Einstein equations, opening the avenue for their observation with future, more precise GW detectors. The intermediate behaviour of tails should, therefore, be understood as a ``long slowly-decaying transient''~\cite{DeAmicis:2024eoy}, possibly making them prone to interacting with structures with a much larger length scale than the remnant BH radius, such as galactic distributions of matter. In fact, these astrophysical environments are known to affect the low-frequency content of the binary coalescence during the inspiral~\cite{Cardoso:2021wlq, Cardoso:2022whc, Duque:2023seg, Vicente:2025gsg, Tomaselli:2024bdd, Tomaselli:2024dbw}, but have much less impact on the high-frequency content of the ringdown~\cite{Spieksma:2024voy}.

The same discussion applies to GW memory. This effect corresponds to a net displacement of free-falling observers by the passage of a GW (see \cite{Compere:2019qed, Mitman:2024uss} for recent reviews). Memory can be divided into two types: ordinary and null (previously labeled as linear and non-linear memory, respectively). The former is related to changes in the energy content of the source of GWs~\cite{1974SvA....18...17Z} and the latter to the backreaction on the spacetime of travelling GWs. The spectrum associated to memory goes like $1/f$ (where $f$ is the frequency), again making it prone to interact with astrophysical environments at low frequencies (large wavelengths). Once again, they have been recently observed for the first time in Numerical Relativity simulations~\cite{Mitman:2024uss}. Another important recent result was that echoes in horizonless compact objects leave a distinct imprint in gravitational memory that optimizes the detection of these exotic objects~\cite{Deppe:2025pvd}.

The main purpose of this study is to, therefore, test a possible dependence of late-time tails and (linear) memory on astrophysical environments in a simplified scenario and draw consequences for GW astronomy. Unless stated otherwise, we adopt geometrized units with $G=c=1$.

\section{Setup}
Only a handful of exact solutions describing the spacetime of matter distributions around BHs are known, preventing a systematic study of the effect of astrophysical environments to late-time tails and memory. We focus on a solution which has been extensively used in several studies as a proxy for generic galactic environments~\cite{Cardoso:2021wlq, Cardoso:2022whc, Speeney:2024mas, Spieksma:2024voy, Pezzella:2024tkf}, corresponding to a spherically symmetric BH surrounded by an anisotropic fluid representing the (dark matter) halo of a galaxy~\cite{Cardoso:2021wlq}. Its line element is
\be
ds^{2}=-f(r) dt^{2}+\frac{dr^{2}}{g(r)}+r^{2}d\theta^{2}+r^{2}\sin^{2}\theta d\varphi^{2}\,,\label{geometry}
\ee
with
\be
f(r)=\left(1-\frac{2M_{\rm BH}}{r}\right)e^\Upsilon \quad , \quad g(r)=1-\frac{2m(r)}{r} \, ,\label{a_dark_matter}
\ee
where the redshift factor
\be
\Upsilon=\sqrt{\frac{M_{\rm H}}{\xi}}\left(2\arctan\left(\frac{r+a_{\rm H}+M_{\rm H}}{\sqrt{M_{\rm H}\xi}}\right)-\pi\right),
\ee
with $\xi=2a_{\rm H}-M_{\rm H}+4M_{\rm BH}$, and we prescribe the mass function
\begin{align}
m(r)&=M_{\rm BH}+\frac{M_{\rm H}r^{2}}{(a_{\rm H}+r)^{2}}\left(1-\frac{2M_{\rm BH}}{r}\right)^{2}.\label{mass-density}
\end{align}
This geometry and matter profile describe
a halo of matter of mass $M_{\rm H}$ and typical length $a_{\rm H}$ surrounding a non-spinning BH of mass $M_{\rm BH}$~\cite{Cardoso:2021wlq}. We can define its compactness 
\begin{equation}
\mathcal{C} = \frac{M_\text{H}}{a_\text{H}}\,.
\end{equation}
For more details on this background spacetime, we refer the reader to Refs.~\cite{Cardoso:2021wlq, Cardoso:2022whc, Spieksma:2024voy}, which studied the impact of this structure on the evolution of EMRIs and on the prompt ringdown of a massive BH coalescence. Other works have extended this construction to more general density profiles~\cite{Speeney:2024mas, Figueiredo:2023gas}, but the qualitative results we will obtain are the same, independently of the density profile and mass function prescribed. For reference, the Milky Way has Sagittarius A* at its center with $M_\text{BH} \sim 4 \times 10^6 M_\odot$, and a halo with $M_\text{H} \sim 10^{12} \, M_\odot$ and $a_\text{H} \sim 2 \times 10^2 \, \text{kpc} \sim 10^{13} M_\text{BH}$, which yields $\mathcal{C} \sim 10^{-7}$~\cite{2019A&A...621A..56P}.  

GWs couple to the effective fluid describing the matter distribution and their propagation should then be affected, even if the spacetime asymptotics are the same. In order to isolate this effect, we consider a minimally coupled scalar field $\Psi$ in this background, whose dynamics are governed by the Klein-Gordon equation
\be
\square \psi = S \, ,
\ee
where $S$ is some possible source term. We expand the scalar field as
\begin{align}
    \psi\left(t,r,\theta,\varphi\right)&=\sum_{\ell=0}^{\infty}\sum_{m=-\ell}^{\ell}\frac{\Psi_{\ell m}(t, r)}{r}Y_{\ell m}\left(\theta, \varphi\right),
    \label{scalarfieldguess}
\end{align}
where $Y_{\ell m}$ are the standard spherical harmonics, $\ell$ is the orbital number, and $m$ is the magnetic quantum number~\cite{Berti:2009kk}. Using this ansatz in the Klein-Gordon equation
\be
\frac{\partial^{2}\Psi}{\partial r_{*}^{2}}-\frac{\partial^{2}\Psi}{\partial t^{2}}-V\Psi=r f S\,,\label{Wave-equation2}
\ee
where we omit the angular subscripts from now on to avoid cluttering. The tortoise coordinate is 
\be
\frac{dr}{dr_{*}}\equiv\sqrt{fg}\,,
\ee
and the effective potential
\begin{align}
    V=f\frac{\ell (\ell+1)}{r^{2}}+\frac{\left(fg\right)'}{2r}\,,
    \label{eq:RW-potential1}
\end{align}
with a prime standing for a derivative wtih respect to $r$.
\subsection{Numerical methods and initial data}

Given some initial data for $\Psi$ and/or a specific source term, Eq.~\eqref{Wave-equation2} can be numerically integrated with standard techniques. However, comprehensive studies on the late time behavior and memory effect require robust methods to extract the GW signal at the infinitely far wave zone, i.e. at future null infinity $\scri^+$. Therefore, we solve Eq.~\eqref{Wave-equation2} in the context of the hyperboloidal framework~\cite{Zenginoglu:2011jz,PanossoMacedo:2023qzp}, and employ two different hyperboloidal formulations, together with two independent solvers to benchmark our results.

In general, an hyperboloidal foliation $\tau$ with compact radial coordinate $\rho$ follows from the transformation~\cite{Zenginoglu:2007jw}
\begin{equation}
    t = \tau - h(\rho) \quad , \quad r_* = x(\rho) \, ,
\end{equation}
with $h(\rho)$ the so-called height function, and $x(\rho)$ a function mapping $r\rightarrow \infty$ into a finite coordinate $\rho_{\scri^+}$.

The first formulation introduces a hyperboloidal foliation by concatenating the usual $t=\text{constant}$ hypersuface into a hyperboloidal layer $\tau=\text{constant}$ at some location $\rho_\text{layer}$ and having $\scri^+$ at the coordinate location $\rho_{\scri^+}=\mathcal{S}$ via~\cite{Zenginoglu:2011zz, Zenginoglu:2012us}
\begin{align}
    x(\rho) &= \dfrac{\rho}{\Omega(\rho)}, \quad h(\rho) =\rho -  \dfrac{\rho}{\Omega(\rho)}, \\
    \Omega(\rho) &=  1 - \left( \frac{\rho - \rho_\text{layer}}{\mathcal{S} - \rho_\text{layer}} \right)^4 \Theta \left(\rho - \rho_\text{layer} \right) \, , 
\end{align}  
where $\Theta$ is the Heaviside function.

The resulting wave equation is then solved with a two-step Lax-Wendroff algorithm with second-order finite differences~\cite{Sundararajan:2007jg, Zenginoglu:2012us, Zenginoglu:2011zz}. 
Even though this strategy provides access to future null infinity, it does not include the BH horizon at $r_* \rightarrow -\infty$ in the numerical grid.

A second formulation -- used in an independent numerical code to cross-check our findings -- employs a hyperboloidal foliation parametrising the entire exterior region via the minimal gauge~\cite{PanossoMacedo:2023qzp}
\begin{eqnarray}
    x(\rho) &=& 2M_{\rm BH}\left(  \dfrac{1}{\rho} - \ln \rho + \ln(1-\rho) \right),\\
    h(\rho) &=& 2M_{\rm BH}\left(  -\dfrac{1}{\rho} + \ln\rho + \ln(1-\rho) \right). 
\end{eqnarray}
In this case, the BH horizon is at $\rho_{\rm H} = 1$, whereas future null infinity at $\rho_{\scri^+} = 0$. The wave equation is then solved with a multi-domain, fully spectral code, with a numerical approximation based on the Chyebyshev polynomials for both the spatial and time directions~\cite{PanossoMacedo:2014dnr}.

When studying the homogeneous problem ($S = 0$), we prescribe as initial data
\begin{align}
\Psi(t=0, r^*) & = \exp \left[-\frac{(r^* - r_0)^2}{2 \lambda^2} \right] / \left( \sqrt{2 \pi} \lambda \right) \, , \label{eq:ID1}\\
\frac{\partial \Psi}{\partial t}(t=0, r^*) & = -2\frac{\left(r^* - r_0 \right)}{\lambda^2} \Psi(t=0, r^*) \, , \label{eq:ID2}
\end{align}
corresponding to an ingoing Gaussian of width $\lambda$, localized around some radius $r_0$. From now on, we work in units where $M_\text{BH} = 1$. Data available in Ref.~\cite{CoG_v2}.

\section{Results}
%
\begin{figure*}[hbt!]
    \centering \includegraphics[width=0.95\columnwidth]{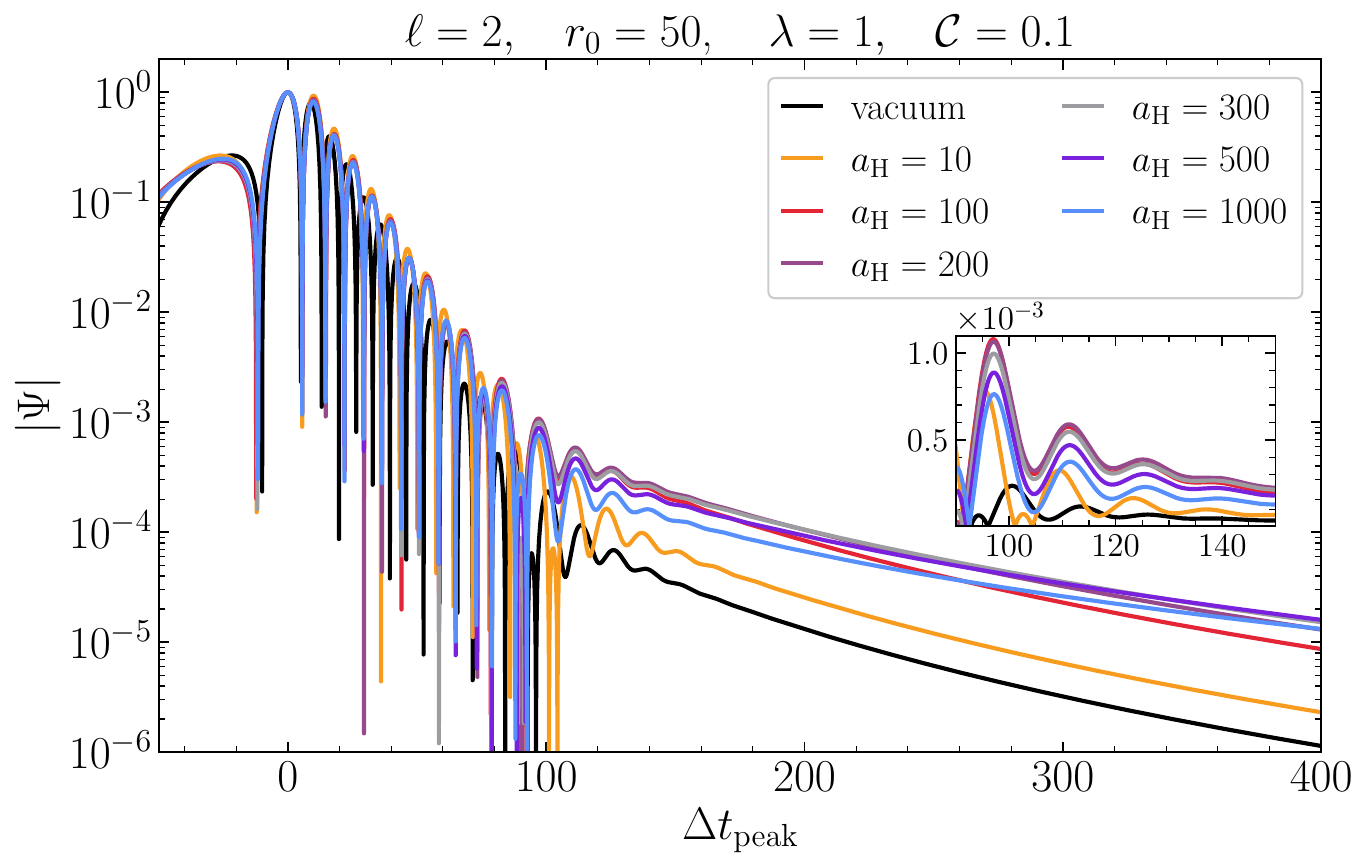} 
     \includegraphics[width=0.95\columnwidth]{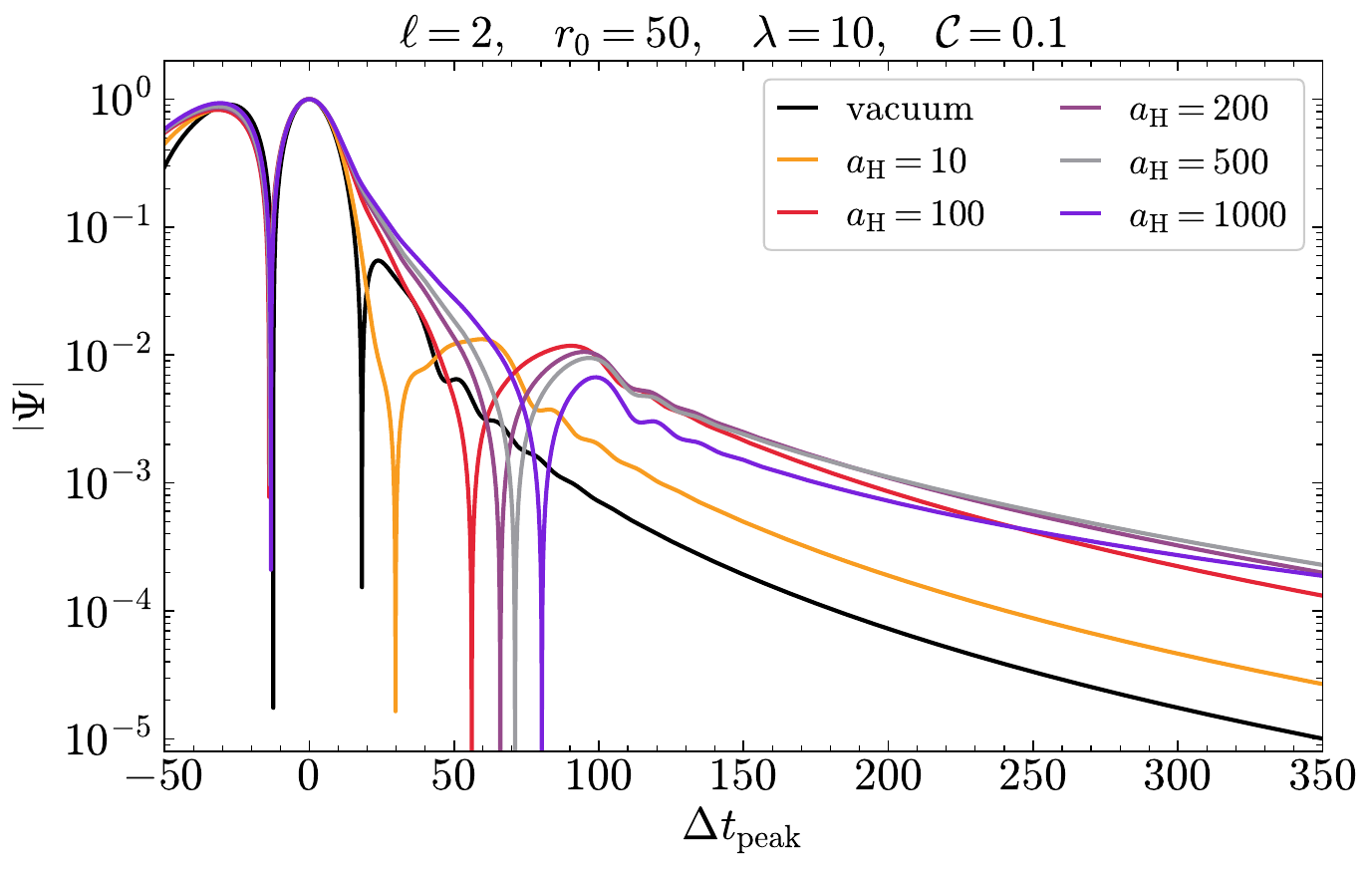}\\
     \includegraphics[width=0.95\columnwidth]{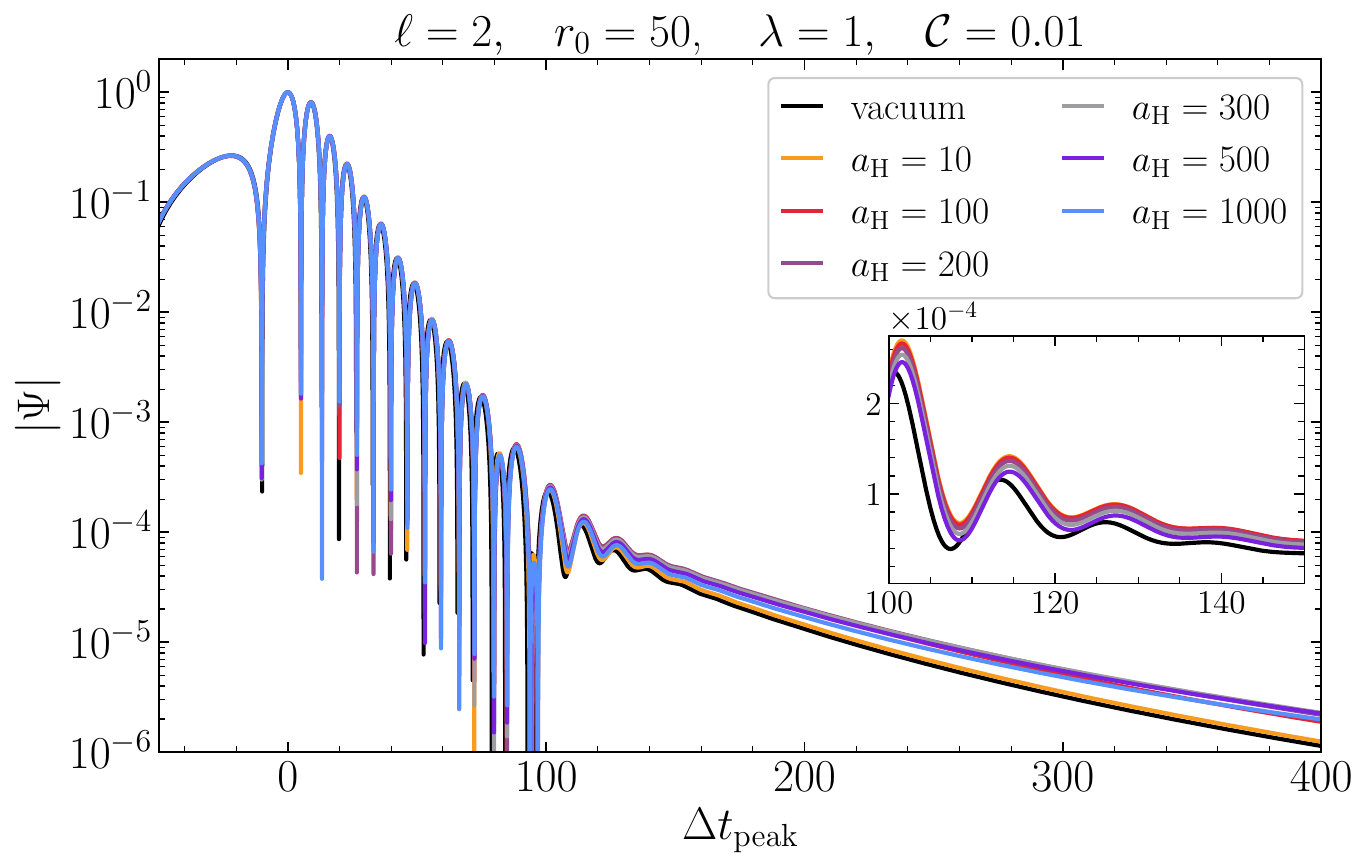} 
     \includegraphics[width=0.95\columnwidth]{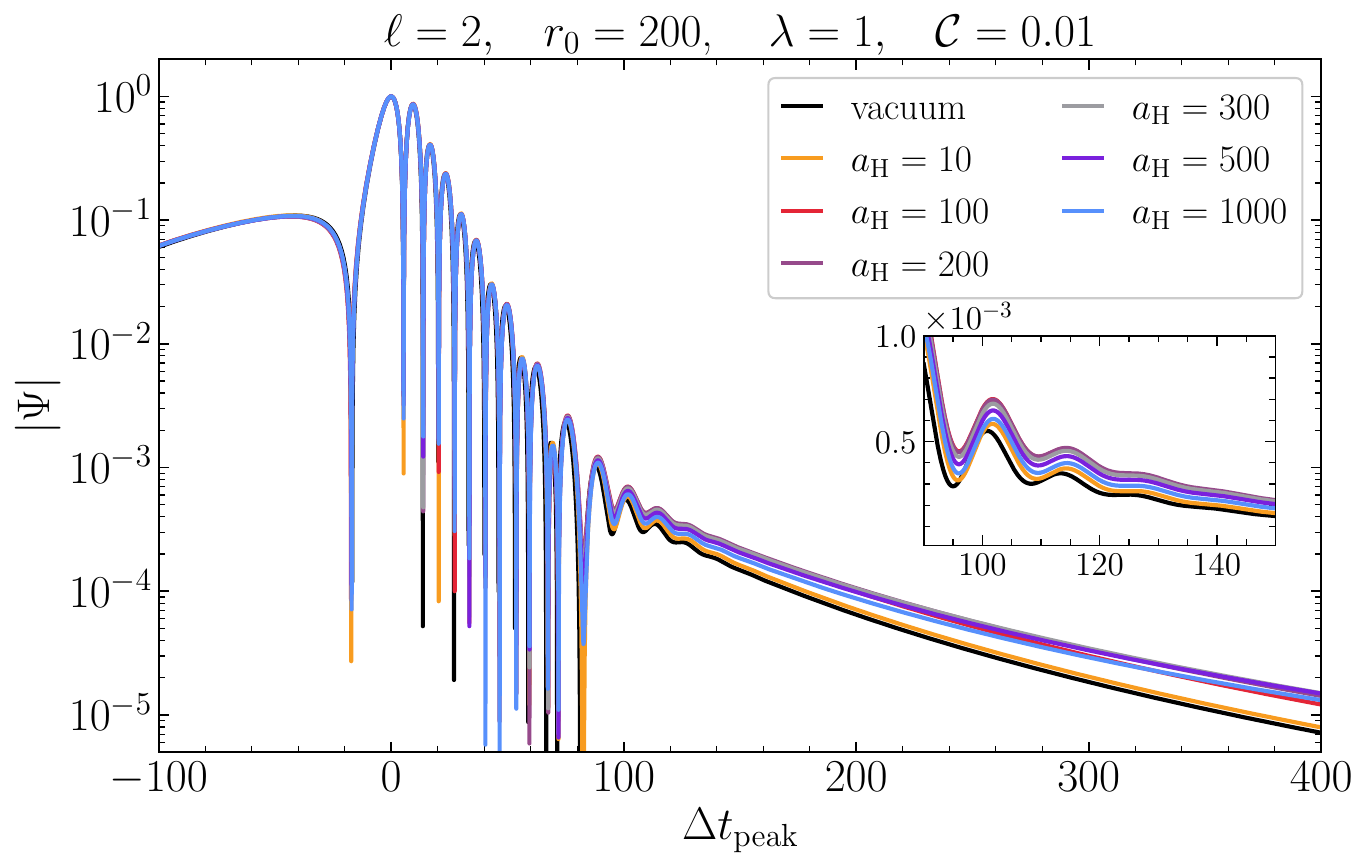}\\
     \includegraphics[width=0.95\columnwidth]{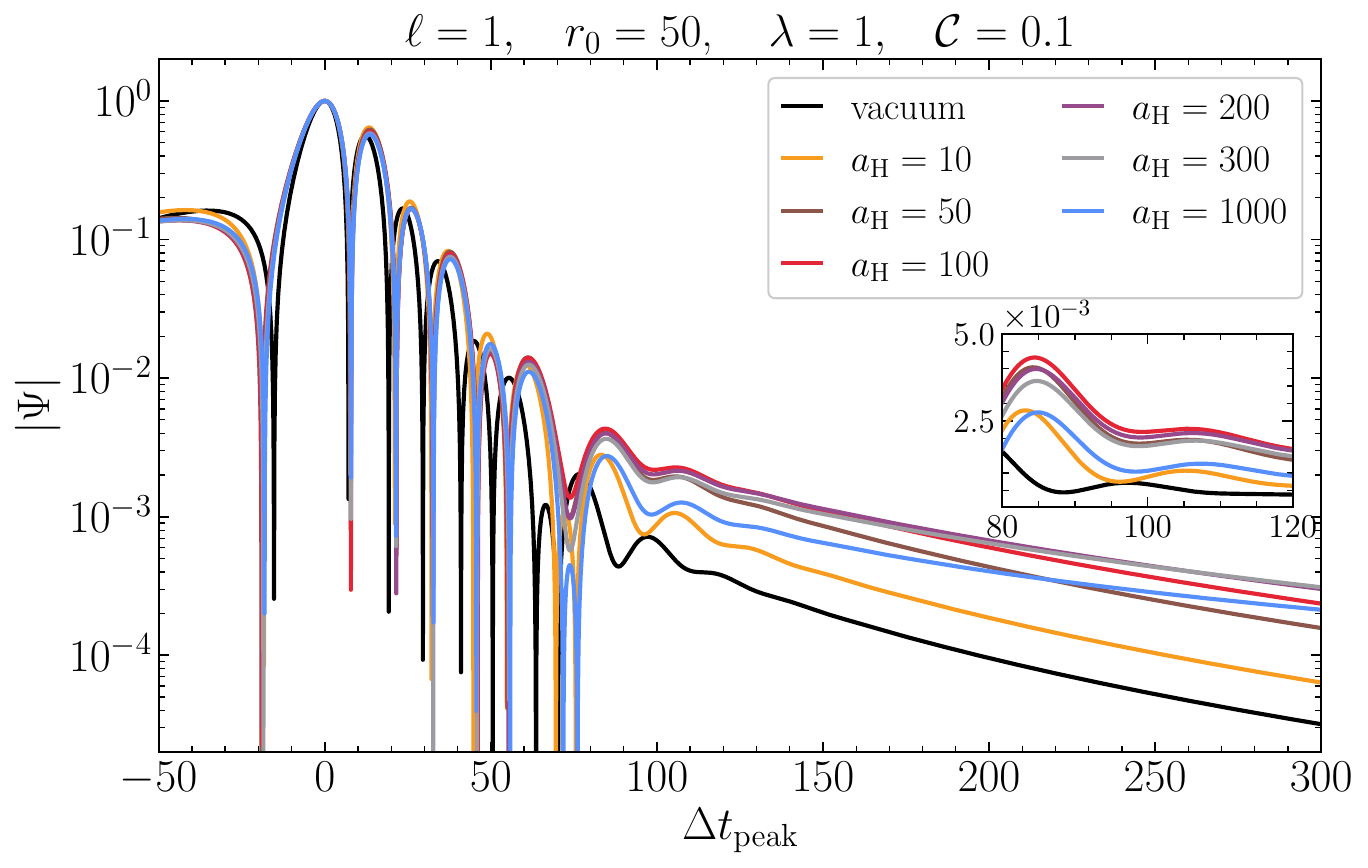} 
     \includegraphics[width=0.95\columnwidth]{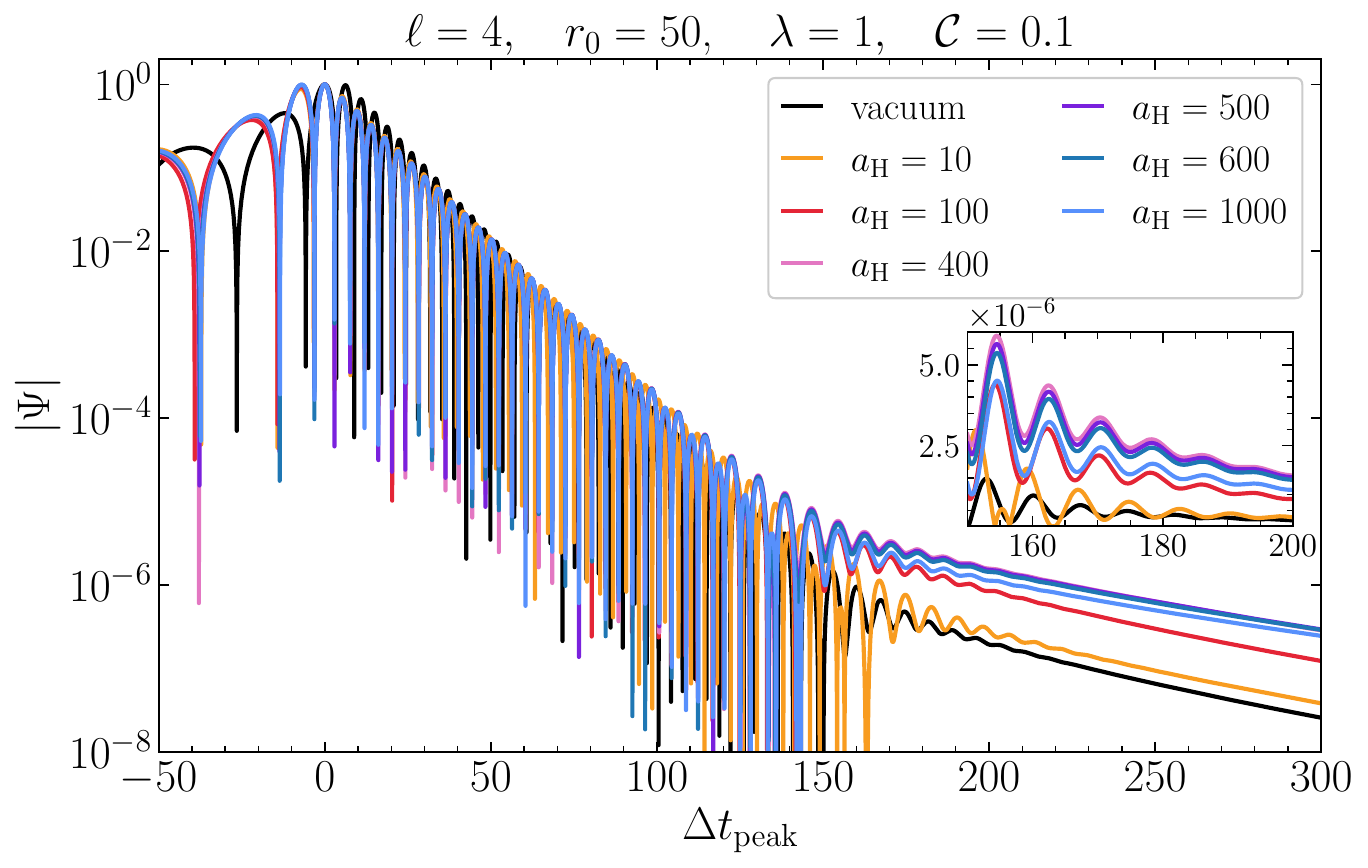}
    \caption{Late-time tails for ingoing Gaussian initial data as prescribed in Eqs.~\eqref{eq:ID1}-\eqref{eq:ID2}. We show results for different halo configurations/compactnesses, gaussian radii $r_0$ and width $\lambda$, and multipoles. Here and in all plots, results are in units of BH mass $M_{\rm BH}$. At very late times, the power-law decay is identical to vacuum, given by Price's dominant decay, but that the transience behavior and amplitude of tails depends on the environment.}
    \label{fig:spin0GaussianplotsF}
\end{figure*}
%
\subsection{Impact of astrophysical environments on initial-data driven tails}

We start with initial-data driven tails, by studying the dependence on halo properties of the signal that ensues after the QNM ringdown. One can show that any effective potential as in Eq.~\eqref{eq:RW-potential1} that differs asymptotically from vacuum by terms $1/r^{\alpha}$, with $\alpha \geq 3$ has the same asymptotic power-law decay as in vacuum~\cite{Ching:1995tj,Ching:1994bd,Cardoso:2024jme,Rosato:2025rtr}. As we make explicit below, the halo solution we focused on also has this behavior. Yet, as stressed in the Introduction, how this asymptotic value is approached, i.e. the intermediate behavior, is highly sensitive to the GW source history and potentially to the interaction of GWs with matter as they propagate. Since it is the intermediate behavior that carries larger power, we focus on this transition regime here. To test this, we evolve initial data as described in Eqs.~\eqref{eq:ID1}-\eqref{eq:ID2} for different initial radii, widths, multipoles, and halo configurations. 

Our results are summarized in Fig.~\ref{fig:spin0GaussianplotsF}, where we normalize $\Psi$ and align the time coordinate by the maximum at the prompt ringdown. Even if asymptotically at late times the decay of perturbations follows the same power-law as in vacuum~\cite{Ching:1995tj,Ching:1994bd,Cardoso:2024jme,Rosato:2025rtr}, there is a clear dependence on the environment in the initial tail that starts dominating the signal. The most striking is the enhancement of the tail amplitude, which for halo compactnesses of $\mathcal{C}=0.1$ can be at least one 1 order of magnitude larger than in vacuum (e.g. top left panel). Lowering the compactness (e.g. middle left panel), weakens the enhancement but the hierarchy between which halo size $a_\text{H}$ optimizes tail enhancement is maintained. Looking at the inset on the top and middle left panel, the initial amplitude of the tail is maximized for the configuration with $a_\text{H} \sim 100-200$. This behavior appears to be independent of the compactness, and initial radius and width of the Gaussian pulse (top and middle right panel). However, as seen in the bottom panels, it does depend on the multipole $\ell$. For $\ell=1$, the $a_\text{H} = 100$ configuration clearly has the tail with the largest initial amplitude while for $\ell = 4$, with the same initial radius, width, and halo compactness, it is $a_\text{H} = 400$ that maximizes tail excitation.  We do not have a concrete explanation for this behavior, but hypothesize it is due to constructive/destructive interference of backscattered waves emitted during the ringdown, whose wavelength depends on $\ell$, that is fine tuned for some halo length scale. We leave a further exploration of this feature with semi-analytic methods for future work~\cite{DeAmicis:2024not}.

\begin{figure*}[hbt!]
\centering 
     \includegraphics[width=0.95\columnwidth]{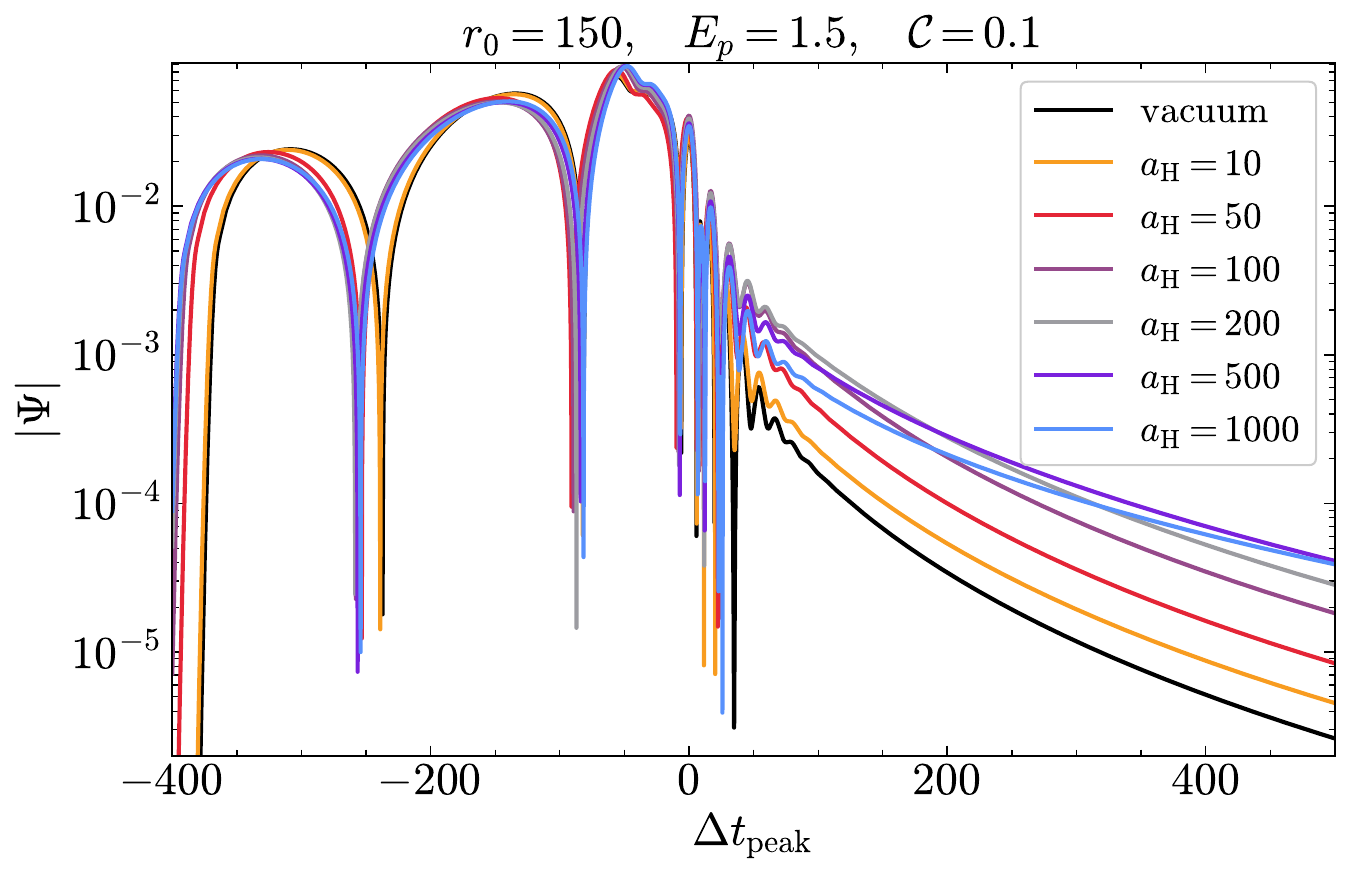} 
     \includegraphics[width=0.95\columnwidth]{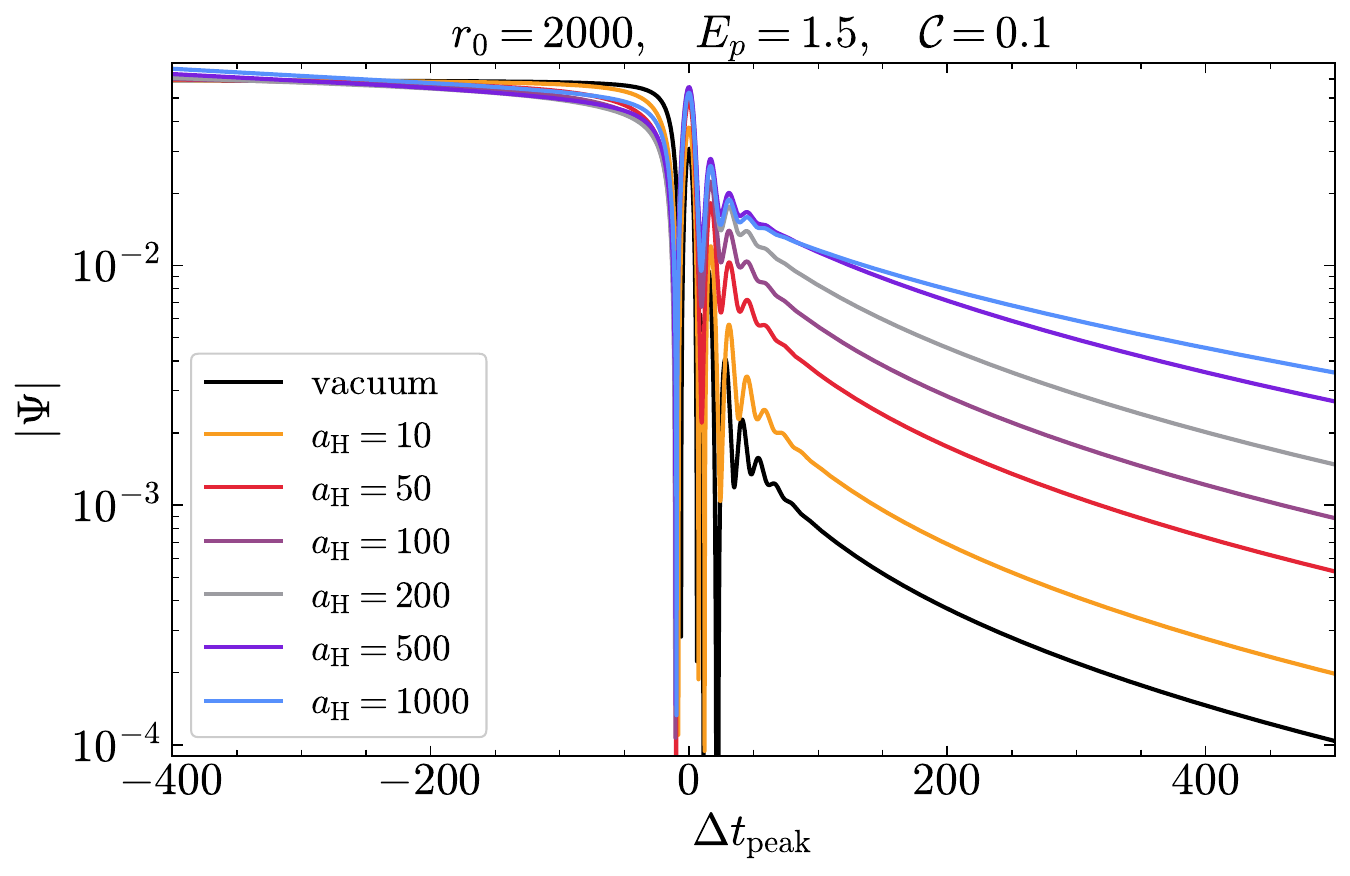}\\
     \includegraphics[width=0.95\columnwidth]{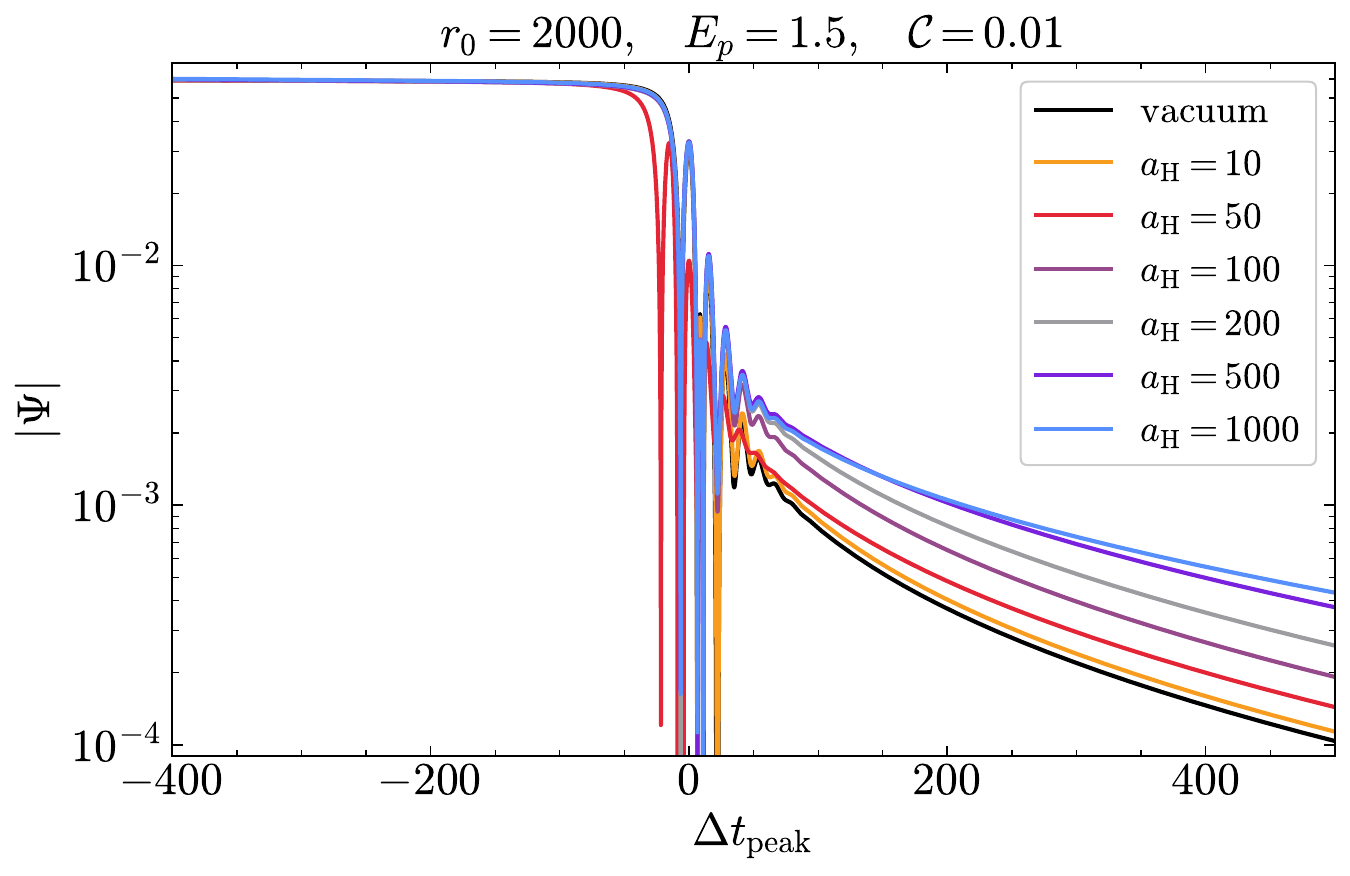}
     \includegraphics[width=0.95\columnwidth]{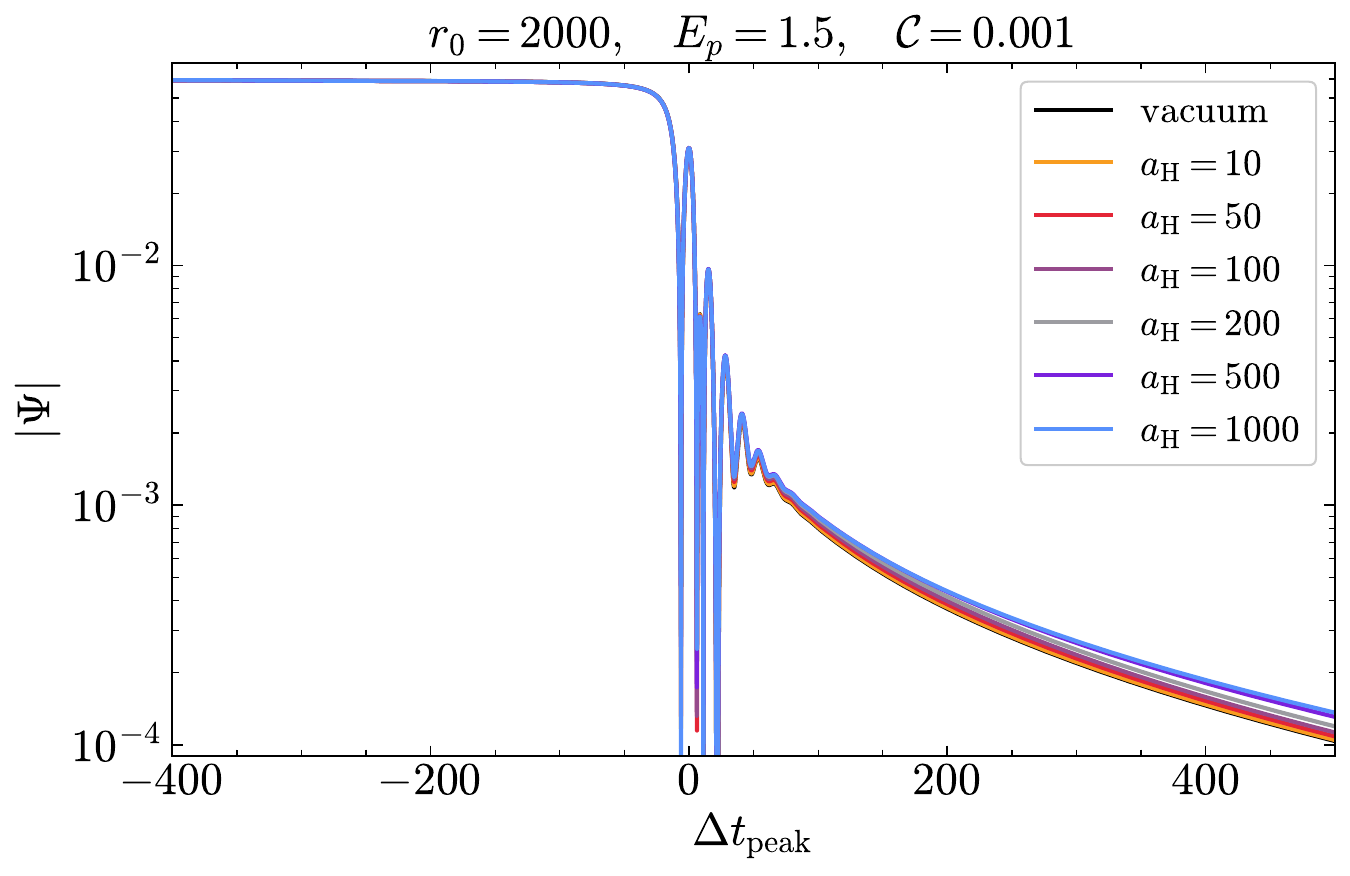}\\
     \includegraphics[width=0.95\columnwidth]{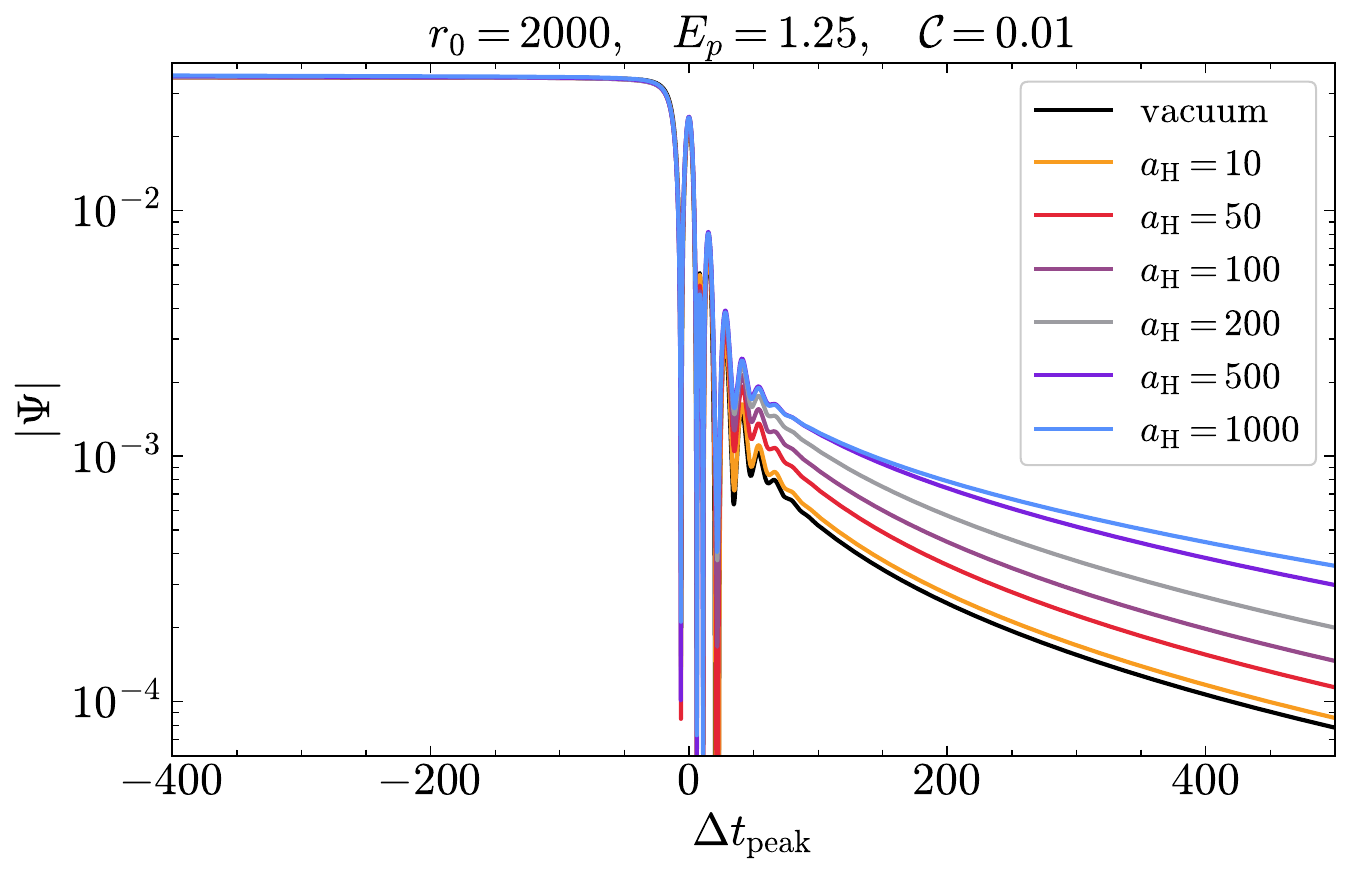}
     \includegraphics[width=0.95\columnwidth]{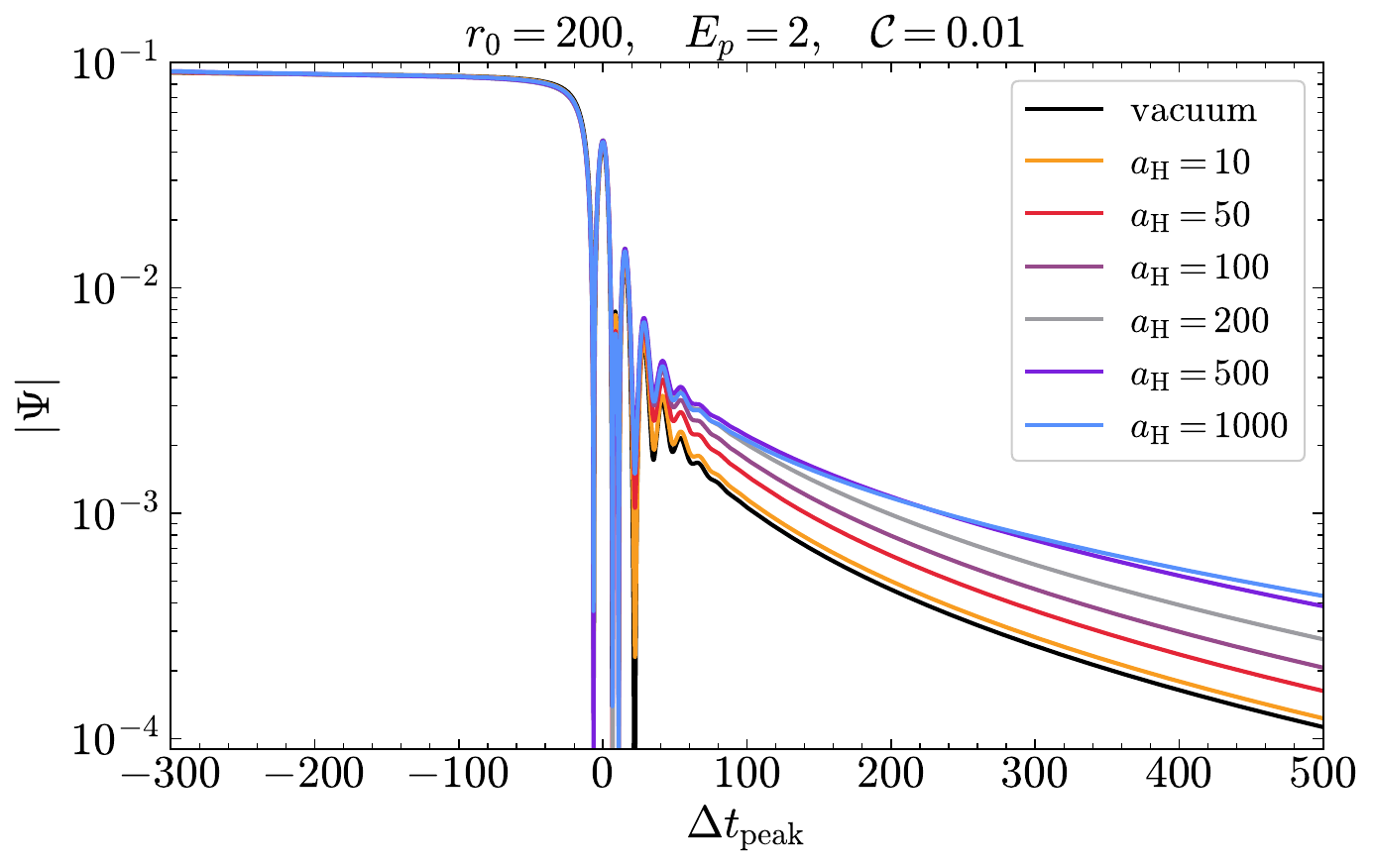}
\caption{Radial infall of a particle in a BH surrounded by a galactic halo, for different halo configurations, initial radius, and energy of the point particle. At very late times, the power-law decay is identical to vacuum, given by Price's dominant decay, but that the transience behavior and amplitude of tails depends on the environment. It is interesting to see that the memory (the difference between the amplitude asymptotically early and late) is independent of environment, but the intermediate transient behavior does depend on the halo properties.}
\label{fig:LinearMemory}
\end{figure*}
Another feature present is that larger halo masses slow down the initial tail decay, i.e. decrease the initial power-law exponent in $\Psi \sim t^{-p}$. This behavior is expected just from expanding the effective potential in Eq.~\eqref{eq:RW-potential1} to next to leading order as $r \rightarrow \infty$
\be
 V = \frac{\ell(\ell+1)}{r^2} + 2(M_\text{BH}+M_\text{H}) \frac{1 - \ell(\ell+1)}{r^3} + \mathcal{O}\left(\frac{1}{r^4}\right) \, .   
\ee
As anticipated, the leading order term in $1/r^2$ is the same as in vacuum, and therefore the asymptotic power-law decay at very late times is also the same~\cite{Ching:1995tj,Ching:1994bd,Rosato:2025rtr} . However, the correction in $1/r^3$ grows linearly with $M_\text{H}$ and, therefore, for larger halo masses the tail takes more time to reach the vacuum power-law tail, implying they dominate asymptotically, i.e.~have larger amplitudes at late times, compared to lower mass distributions. Indeed, we have performed very long and stable time evolutions and observed that the larger the halo mass, the longer it gets to achieve the expected asymptotic power-law decay rate.

\subsection{Impact of astrophysical environments on the linear memory and source driven tails}

We now move to the non-homogeneous problem and study the impact of the galactic halo on source driven tails and (linear) memory. As should be clear from our framework, we investigate only linear, scalar tails and memory driven by a scalar charge. We expect the gravitational case to display similar features. Power-law decay tails are driven by the back-scattering of spacetime curvature. They are universal (present for all massless fields) features of curved spacetime and have been investigated at length~\cite{Price:1971fb,Gundlach:1993tp,Ching:1995tj,Blanchet:1994ez,DeAmicis:2024eoy, Cardoso:2024jme}. Linear gravitational memory effects are triggered by the asymptotic states, in particular by the fact that initial and end states have different asymptotic velocities~\cite{1974SvA....18...17Z,Adler:1975dj,Cardoso:2003cn}. Their dependence on the velocity of the incoming particle is different for scalar and gravitational waves, a flat-space description of these effects can be found in Refs.~\cite{Smarr:1977fy,Berti:2010ce,GRIT1,GRIT2}. We provide in Appendix~\ref{Appendix} a description of the memory for scalar charges. In particular, infalls from rest produce linear gravitational memory after plunging onto a BH, due to no-hair properties of BHs. However, the linear memory of scalars is of similar nature to that of gravity, despite the different dependence of source on the velocity.

Consider therefore a point particle following radial geodesic infall with some initial velocity, such that
\begin{align}
    \theta_p = \varphi_p = 0 \, , \quad
    \frac{d r_p^*}{dt} = - \sqrt{1 - f(r_p)/E_p^2} \, , 
\end{align}
where $E_p$ is the energy per unit rest mass that characterizes the geodesic and $r_p \equiv r_p(t)$ is the radius of the particle at each instant in time. For the functional form of the source term $S$ in the wave equation~\eqref{Wave-equation2}, we take the trace of the energy-momentum tensor of the point particle projected onto each harmonic, such that
\begin{align}
S &= \int d\Omega \,  Y^*_{\ell m} (\theta,\varphi)  \,  (T^p)^\mu_\mu  \, , \\
T_p^{\mu \nu} &= \frac{q_p}{r^2 \sin \theta }\frac{E_p}{f}\frac{d x^\mu}{dt}\frac{d x^\nu}{dt} \delta\left[r-r_p(t)\right] \delta\left[\theta-\theta_p(t)\right] \nonumber \\ & \times  \delta\left[\varphi-\varphi_p(t)\right]  \, . 
\end{align}
The Dirac delta in the radial direction is approximated by a narrow width Gaussian as described in Refs.~\cite{Lopez-Aleman:2003sib, Zenginoglu:2011zz}. Here $q_p$ is the scalar charge of the infalling particle. Our results scale trivialy with $q_p$ and we set it to unit onwards.

Our results are illustrated in Fig.~\ref{fig:LinearMemory} for different halo configurations, initial radius and proper energy of the particle. The first thing to remark is that the $r_0 = 150$ (top left panel) case does not show the same initial behavior as when the initial radius is $r_0 = 2000$, i.e. the constant plateau which actually corresponds to the linear memory. In other words, the initial radius has to be sufficiently large so that `true' physical signal builds up to dominate over the junk of radiation due to the prescription of non-consistent initial data (this is what is observed for the $r_0=150$ case before the ringdown).
Note that even if the initial radius is very large, for larger masses and compactnesses ($\mathcal{C} = 0.1$), the memory effect appears to not have fully converged (i.e. the initial plateau is not horizontal). However, for smaller compactnesses, describing more realistic distributions of matter at astrophysical scales, the memory waveform is closer to the corresponding vacuum one, indicating that astrophysical environments do not impact linear memory significantly.

More interesting is the behavior of the tails. For small initial radius ($r_0 = 150$), where for some configurations the particle starts the motion already deep inside the halo, the behavior is similar to the previous section, i.e. halos with typical size $a_\text{H} \sim 100-200$ optimize tail excitation. However, when the particle starts outside the halo, the initial amplitude of the tail is larger for larger halo masses. For the same configurations, smaller energies (corresponding to smaller initial velocities) promote tail enhancement with respect to vacuum, suggesting that the more time the particle spends in the halo, and is accelerated by it, the larger the tail. This aspect of our numerical results agrees with the analytical expansions in Ref.~\cite{DeAmicis:2024not}. Note also that, for the same halo configurations, the relative amplitude of the tail with respect to the vacuum case is much larger than in the homogeneous problem studied in the previous section. The upshot is that while (linear) memory appears insensitive to distributions of matter around the central BH, tails are much more effective at probing the environment.

\section{Discussion}

The missing piece in our study is to connect the phenomenological results of our simplified toy model with astrophysical systems. Considering the typical compactness of galaxies ($\mathcal{C} \sim 10^{-6} - 10^{-7}$), it is clear from our results for the lowest compactnesses that the effects observed for tail enhancement are completely negligible for realistic dark matter halos.

However, could a more ``exotic'' environment reach the compactnesses probed in our numerical experiments? Recent numerical relativity simulations of BHs merging in an initially homogeneous profile of wave dark matter have shown that overdensities are formed around the two BHs up until merger~\cite{Aurrekoetxea:2023jwk, Aurrekoetxea:2024cqd}. Reading off from simulations, e.g. Fig.~4 in \cite{Aurrekoetxea:2024cqd} for $\lambda$ (no self-interactions of the scalar field), at $t=1600\, M_\text{BH}$, the central region ($\lesssim 5-10 M_\text{BH}$) has a density of $\rho \sim 10^3 \rho_0$ where $\rho_0$ is the initial density of the homegeneous environment in code units. Assuming that the distribution in this region is spherically uniform, then the mass of the environment is $M_\text{H}(r=10M_\text{BH}) \approx 4\times10^6 \rho_0$, and its compactness is $\mathcal{C} \sim  4\times10^6 \rho_0 / 10 \sim 4\times 10^5\rho_0$. The only mechanism we envision capable of generating $\rho_0$ large enough so that $\mathcal{C} \gtrsim 5\times 10^{-3}$ is via BH superradiance. It would therefore be important to understand how these structures evolve during a comparable mass binary coalescence and study not only the dephasing but also the post-merger signal, in particular the tail structure. 

Despite its potential astrophysical (non-)relevance our study shows that indeed there is an impact of environments to the intermediate behavior of tails and one needs to be careful when extrapolating asymptotic behaviors, as the ones obtained in Ref.~\cite{Rosato:2025rtr}, to the transient part of the signal.

\begin{acknowledgments}
We thank the anonymous referee for useful suggestions and comments.
We thank Keefe Mitman and Jaime Redondo-Yuste for fruitful discussions and clarification on the memory effect. We are indebted to Gregorio Carullo, Marina de Amicis, Paolo Pani and Romeo Rosato for useful feedback and comments on the final manuscript. The Center of Gravity is a Center of Excellence funded by the Danish National Research Foundation under grant No. 184.
We acknowledge support by VILLUM Foundation (grant no. VIL37766) and the DNRF Chair program (grant no. DNRF162) by the Danish National Research Foundation.
V.C.\ is a Villum Investigator and a DNRF Chair.  
V.C. acknowledges financial support provided under the European Union’s H2020 ERC Advanced Grant “Black holes: gravitational engines of discovery” grant agreement no. Gravitas–101052587. 
Views and opinions expressed are however those of the author only and do not necessarily reflect those of the European Union or the European Research Council. Neither the European Union nor the granting authority can be held responsible for them.
This project has received funding from the European Union's Horizon 2020 research and innovation programme under the Marie Sklodowska-Curie grant agreement No 101007855 and No 101131233.
The Tycho supercomputer hosted at the SCIENCE HPC center at the University of Copenhagen was used for supporting this work.
\end{acknowledgments}

\appendix

\section{On the scalar memory and low-frequency results in flat space\label{Appendix}}
Here we revisit a classical calculation, in the gravitational sector, 
related to the instantaneous collision of two point particles in flat space~\cite{Smarr:1977fy,Adler:1975dj,Berti:2010ce,GRIT1,GRIT2}. This model reproduces extremely well the plunging of a particle onto a massive BH~\cite{Cardoso:2002ay,Berti:2010ce} and even the full nonlinear, high-energy collision of two BHs~\cite{Sperhake:2008ga}. Here we adapt it to the collision of scalar charges, generalizing the results of Ref.~\cite{GRIT1,GRIT2}.

Take a pointlike scalar charge $q_p$ freely moving in flat space along the $z-$axis with four-velocity $v^\mu$ towards a massive obstacle, where it stops. At $t=0$ the two objects collide, and one is left with a single object with charge $q_p\delta$. Here, we allow for charge conservation ($\delta=1$) or charge disappearance ($\delta=0$), the former to mimic the collision with BH (motivated by no-hair results, BHs carry no scalar charge).
We have the stress tensor
\be
T^{\mu\nu}=q_p\frac{v^\mu v^\nu}{\gamma}\delta^{(3)}(\vec{x}-\vec{v}t)\Theta(-t)+q_p\delta\frac{v'^\mu v'^\nu}{\gamma'}\delta^{(3)}(\vec{x})\Theta(t)\,,
\ee
with $\gamma=1/\sqrt{1-v^2}$ and primed quantities stand for the final object at rest (hence $\gamma'=1$). The collision is instantaneous so there is obviously a price to pay for this approximation. It fails to take into account the structure of the colliding particles and the timescales of the collision process. Fortunately, memory effects are oblivious to these scales. The Fourier transform $\tilde{T}$ yields
\be
\tilde{T}^{\mu\nu}=q_p\left(\frac{v^\mu v^\nu}{2\pi i\gamma (\omega-\omega v\cos\theta)}-\delta \frac{v'^\mu v'^\nu}{2\pi i \omega}\right)\,,
\ee
where $\theta$ is the angle between the particle motion direction and the direction of observation of the scalar wave. The scalar flux per unit solid angle 
\be
\frac{dE}{d\omega d\Omega}=\frac{\omega^2}{4}\left|  T^\mu_\mu\right|^2=\frac{q_p^2}{16\pi^2}\left(\frac{1-\delta\gamma(1-v\cos\theta)}{\gamma (1-v\cos\theta)}\right)^2\,.
\ee

There are two important aspects in this result, all consequence of a flat (frequency-independent) spectrum: at large frequencies the integrated energy diverges, a divergence which can only be cured by the introduction of a cutoff~\cite{Smarr:1977fy,Cardoso:2002ay,Sperhake:2008ga,Berti:2010ce,GRIT1,GRIT2}.

Secondly, at low frequencies the spectrum is also flat, for any $\delta$. Indeed, for small velocities,
\be
\frac{dE}{d\omega d\Omega}=\frac{q_p^2}{16\pi^2}\left(1-\delta+\cos\theta (2-\delta)v+{\cal O}(v^2)\right)\,.
\ee
Thus, for a BH with $\delta=0$ the spectrum is flat and finite at $\omega=0$ even for arbitrarily small velocities. For $\delta=1$ however, one recovers the gravitational memory feature that memory requires a finite incoming velocity.

We discussed fluxes, but note that the Fourier transform of the time derivative of $\psi$ (we call it $\dot{\bar{\psi}}$)
\be
\dot{\bar{\psi}}_{\omega=0}=\lim_{\omega \to 0} \int^{+\infty}_{-\infty} \dot{\psi}e^{i\omega t}dt=\psi(t=+\infty)-\psi(t=-\infty)\,.
\ee
And because $dE/d\omega d\Omega\propto \left|\dot{\bar{\psi}}\right|^2$, a flat flux at zero frequency implies a memory in the signal itself~\cite{Kovacs:1978eu}.

\bibliography{References}

\end{document}